\documentclass[aps,nofootinbib,preprint,tightenlines]{revtex4}

\usepackage{graphicx}
\usepackage{amsmath}
\usepackage{bm} 

\newcommand{\beq}{\begin{eqnarray}}
\newcommand{\eeq}{\end{eqnarray}}
\newcommand{\bqa}{\begin{eqnarray}}
\newcommand{\eqa}{\end{eqnarray}}

\def\vec#1{{\bf #1}}

\def\mqo2{{\!\!\!}}

\begin{document}

%%%%%%%%%%%%%%%%%%%%%%%%%%%%%%%%%%%%%%%%
%The normal things to start with.
%%%%%%%%%%%%%%%%%%%%%%%%%%%%%%%%%%%%%%%%

\preprint{HISKP-TH-09/15}
\title{On the scattering of $D$ and $D^*$ mesons off the $X(3872)$}
\author{David L. Canham}\email{canham@itkp.uni-bonn.de}
\author{H.-W. Hammer}\email{hammer@itkp.uni-bonn.de}
\affiliation{Helmholtz-Institut f\"ur Strahlen- und Kernphysik (Theorie)\\
and Bethe Center for Theoretical Physics,
 Universit\"at Bonn, 53115 Bonn, Germany}
\author{Roxanne P. Springer}\email{rps@phy.duke.edu}
\affiliation{Department of Physics, Box 90305, Duke University,
Durham, NC, 27708, USA\\}
\date{\today}

%%%%%%%%%%%%%%%%%%%%%%%%%%%%%%%%%%%%%%%%%
%The Abstact
%%%%%%%%%%%%%%%%%%%%%%%%%%%%%%%%%%%%%%%%%

\begin{abstract}
Both the mass (just below the $D^{*0}\bar{D}^0$ threshold)
and the likely quantum numbers ($J^{PC}=1^{++}$) 
of the $X(3872)$ suggest that it is
either a 
weakly-bound hadronic ``molecule'' ($X(3872 \sim 1/\sqrt{2} [
D^{*0}\bar{D}^0 + \bar{D}^{*0}D^0 ]$) 
or a virtual state of charmed mesons. Assuming the $X(3872)$ is
a weakly-bound molecule, the scattering of neutral $D$ and $D^*$ mesons 
off the $X(3872)$ can be predicted from the $X(3872)$ binding energy.
We calculate the phase shifts and cross section for
scattering of $D^0$ and $D^{*0}$ mesons 
and their antiparticles off the 
$X(3872)$ in an effective field theory for short-range interactions.
This provides another example of a three-body process, along with those
in nuclear and atomic systems, that displays universal properties.
It may be possible to extract the scattering  within the 
final state interactions
of $B_c$ decays and/or other LHC events. 

\end{abstract}

\maketitle

%%%%%%%%%%%%%%%%%%%%%%%%%%%%%%%%%%%%%%%%%%%%%%%
%Brief introduction 
%%%%%%%%%%%%%%%%%%%%%%%%%%%%%%%%%%%%%%%%%%%%%%%

\section{Introduction}
\label{sec:Intro}

In recent years many new 
and possibly exotic charmonium states have been observed
 at the 
B-factories at SLAC \cite{Palano:2009zz}, at KEK \cite{Bracko:2009zz}
in Japan, and at the CESR collider at Cornell \cite{Mendez:2007ey}.
This has revived the field of charmonium spectroscopy
\cite{Swanson:2006st,Eichten:2007qx,Voloshin:2007dx,Godfrey:2008nc,
Nielsen:2008tt,Robutti:2009af}, which
will be an important part of PANDA at the FAIR facility \cite{Lange}.
Because several of the new states exist very close to scattering thresholds,
it is useful to interpret them as hadronic molecules, a concept
introduced in Refs.~\cite{Voloshin:1976ap,
De Rujula:1976qd,Tornqvist:1991ks} well before these most recent
experiments.
A particularly interesting example is the 
$X(3872)$, discovered by the Belle collaboration
\cite{Choi:2003ue} in $B^{\pm}\to K^\pm \pi^+ \pi^- J/\psi$ decays
and quickly confirmed by CDF \cite{Acosta:2003zx}, D0
\cite{Abazov:2004kp}, and BaBar \cite{Aubert:2004ns}.
The state has likely quantum numbers $J^{PC}=1^{++}$ and
is very close to the $D^{*0} \bar{D}^0$ threshold.\footnote{Note, however,
that $J^{PC}=2^{-+}$ cannot be experimentally excluded at present.} 
As a consequence, the $X(3872)$ has a resonant 
S-wave coupling to the $D^{*0} \bar{D}^0$ system.
Early examples of discussions of the possible molecular nature of the
$X(3872)$ can be found in Refs.~\cite{Close:2003sg,Pakvasa:2003ea,Wong:2003xk}.
An extensive program  (see a status report in 
Ref.~\cite{Braaten:2008nv}) 
provides predictions for its decay modes based on the assumption that
it is a  $D^{*0} \bar{D^0}$ molecule with even C-parity:
\beq
(D^{*0} \bar{D}^0)_+ \equiv \frac{1}{\sqrt{2}}
\left(D^{*0} \bar{D}^0+D^{0} \bar{D}^{*0}\right)\,.
\label{eq:Xflavor}
\eeq

The measured mass and width of the $X(3872)$ differ significantly in the 
$J/\psi \pi^+ \pi^-$ and $D^{*0} \bar{D}^0$ decay channels. This effect can
be understood from a line shape analysis which
shows that the true mass and width of the $X(3872)$ are measured in the 
$J/\psi \pi^+ \pi^-$ channel because the $D^{*0} \bar{D}^0$ channel is 
contaminated by a threshold enhancement \cite{Braaten:2007dw,Braaten:2007ft,
:2008su,Hanhart:2007yq}.
Using the latest measurements in the  $J/\psi \pi^+ \pi^-$ channel 
\cite{Belle:2008te,Aubert:2008gu,CDF-QWG08}, the mass of the 
$X(3872)$ is \cite{Braaten:eft09}:
\beq
m_X = (3871.55 \pm 0.20) \mbox{ MeV}\,,
\label{eq:MX}
\eeq
which corresponds to an energy relative to the $D^{*0}\bar{D}^0$ threshold
\cite{Cawlfield:2007dw} of
\beq
E_X =(-0.26 \pm 0.41) \mbox{ MeV}\,.
\label{eq:EX}
\eeq
The central value corresponds to a $(D^{*0} \bar{D}^0)_+$
bound state with binding energy
$B_X=0.26$ MeV (but a virtual state cannot be excluded from the
current data in the $J/\psi \pi^+ \pi^-$ and $D^{*0} \bar{D}^0$ channels
\cite{Bugg:2004rk,Hanhart:2007yq,Voloshin:2007hh}). 
The $X(3872)$ is also very narrow,
with a width smaller than 2.3 MeV.

Because the $X(3872)$ is so close to the $D^{*0} \bar{D}^0$ threshold,
it has universal low-energy properties that depend only on its binding 
energy \cite{Braaten:2004rn}. 
Close to threshold, the coupling to charged $D$ mesons can be neglected
because the $D^{*+} \bar{D}^-$ threshold is about 8 MeV higher in energy.
Therefore, the properties of the $X(3872)$ can be described
in a universal EFT with contact interactions only.  This EFT
is a  {\it pionless EFT} because the pion degrees of freedom are 
not dynamical near threshold; they are integrated out and all 
effective interactions are short range.
This EFT is widely used in low-energy nuclear physics
\cite{vanKolck:1999mw,Beane:2000fx,Bedaque:2002mn,Epelbaum:2008ga}.
For an EFT of the $X(3872)$ including explicit pions, see 
Refs.~\cite{Fleming:2007rp,Fleming:2008yn}.
The study of the $X(3872)$  as a $(D^{*0} \bar{D}^0)_+$ molecule
in the pionless EFT was initiated
by Braaten and Kusunoki \cite{Braaten:2003he}. A number of predictions
for production amplitudes \cite{Xproduction},
decays \cite{Xdecay}, formation \cite{AlFiky:2005jd}, and line shapes
\cite{Braaten:2007dw,Braaten:2007ft} within this framework followed.
The interactions of the $X(3872)$ with other hadrons, however, are not known.

In this paper, we extend these studies to three-body processes in the pionless
EFT.  Based on the assumption that the $X(3872)$  is 
an  S-wave $(D^{*0} \bar{D}^0)_+$ molecule,
we provide model-independent predictions for the scattering 
of $D^0$ and $D^{*0}$ 
mesons and their antiparticles off the $X(3872)$ in the pionless EFT.
We will refer to these reactions collectively 
as $D^{(*)}X$ scattering. This reaction may
contribute to the final state interaction in decays of $B_c$ mesons 
into $D$ and $D^*$ mesons, in rare events in $B \bar{B}$ production 
where one of the $B$'s decays into an $X$ and the other one into a 
$D$ or $D^*$ meson, and in prompt events at colliders. 
In the next section we will provide the EFT for $D^{(*)}X$ scattering. 
In section \ref{sec:ResDis}
we will present our results and discuss possible scenarios for
observing this process.

\section{Formalism and Calculation}
\label{sec:Form}

In this section we set up the EFT for $D^{(*)}X$ scattering,
derive the integral equation for the scattering amplitude,
and provide an expression for the total cross section.
The formalism can be taken over from the pionless theory in 
low-energy nuclear physics, but we briefly describe the issues
relevant for $D^{(*)}X$ scattering. For a more detailed discussion
and a bibliography of the original work, see the reviews of 
Refs.~\cite{vanKolck:1999mw,Beane:2000fx,Bedaque:2002mn,Epelbaum:2008ga}.
For the derivation
of the three-body equations, it is convenient to introduce a
non-dynamical auxiliary field $X$ for the $X(3872)$.
The EFT is organized in an expansion around the non-trivial fixed
point of the coupling between $D^0$ and $D^{0*}$ mesons
corresponding to infinite scattering length or, equivalently,
the $X(3872)$ being a threshold bound state. 
The binding momentum, $\gamma$, of the $X(3872)$ is $\sqrt{2\mu_X B_X}$, with 
the reduced mass $\mu_X = {m_{D^0}m_{D^{*0}} /(m_{D^0}+m_{D^{*0}})}$.
So an infinite scattering length corresponds to the $\gamma\equiv 0$ limit.
The EFT expansion is then in powers of 
$\gamma/\Lambda_b$ and $k/\Lambda_b$ where
$k$ is the typical momentum exchange and 
$\Lambda_b$ is the breakdown scale of the pionless EFT.
We will estimate $\Lambda_b$ from one-pion exchange in the discussion of
the errors below. To leading order in this expansion, the
effective Lagrangian for the interaction of the $X(3872)$ with 
neutral $D$ and $D^*$ mesons can be written as:
\bqa
{\cal L} &=& \sum_{j= D^0, D^{*0}, \bar{D}^0, \bar{D}^{*0}}
\psi_j^\dagger \left( 
i\partial_t+\frac{\nabla^2}{2m_j}\right) \psi_j +\Delta X^\dagger X
\nonumber\\
&&-\frac{g}{\sqrt{2}}\left( X^\dagger (\psi_{D^0} \psi_{\bar{D}^{*0}} 
+\psi_{D^{*0}} \psi_{\bar{D}^0})
+ \mbox{ H.c. }\right) +\ldots\,,
\label{eq:lag}
\eqa
where  H.c. denotes the Hermitian conjugate and
the dots indicate higher order terms with more derivatives and/or fields.
The terms with more derivatives are suppressed at low energies.
As shown in \cite{Braaten:2003he},
there is no Efimov effect  \cite{Efimov-70}
in this system, so three-body terms will not contribute 
up to next-to-next-to-leading order (N2LO) in the expansion 
in $\gamma/\Lambda_b$.
Four- and higher-body forces do not contribute in three-body
processes.
The only corrections up to N2LO are effective range contributions and their
inclusion is in principle straightforward  
\cite{Bedaque:1997qi,Bedaque:1998mb,Griesshammer:2004pe}.
However, the effective range for the $X(3872)$ is not known.
We will therefore restrict our calculation to leading order only,
and estimate the size of higher order corrections.
The $D^0$, $D^{*0}$, $\bar{D}^0$, and $\bar{D}^{*0}$
mesons are treated as distinguishable particles;
charge conjugation invariance yields
$m_{D^0}=m_{\bar{D}^0}$ and $m_{D^{*0}}=m_{\bar{D}^{*0}}$.

%%%%%%%%%%%%%%%%%%%%%%%%%%%%%%%%%%%%%%%%%%%%%%%%%%%%%%%%%% 
\begin{figure}[t]
\begin{center}
\includegraphics[width=10cm,angle=0,clip=true]{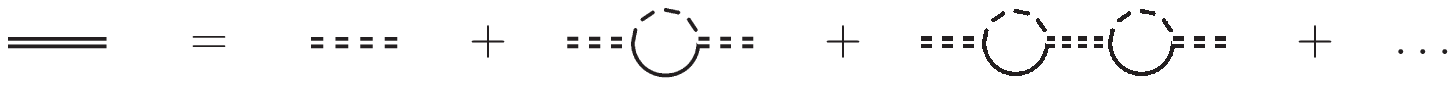}
\end{center}
%\vspace*{-18pt} 
\caption{Dressing of the bare $X$ propagator (double dashed line)
by $D^0$ and $D^{*0}$ meson loops (solid and dashed lines).}
\label{fig:dress}
\end{figure}
%%%%%%%%%%%%%%%%%%%%%%%%%%%%%%%%%%%%%%%%%%%%%%%%%%%%%%%%%%%%%%
The parameters $\Delta$ and $g$ in Eq.~(\ref{eq:lag})
are not independent; only the 
combination $g^2/\Delta$ enters into physical observables. 
Since the theory is nonrelativistic, all particles propagate forward in time
and the tadpoles vanish. The propagator for
the $D^{(*)}$ mesons is 
\begin{equation}
\label{nucprop}
iS_j (p_0,\vec{p})=\frac{i}{p_0-p^{2}/(2m_j) +i\epsilon}\,,
\qquad j= D^0, D^{*0}, \bar{D}^0, \bar{D}^{*0}\,,
\end{equation}
where $p^2 \equiv \vec{p}^{\,2}$.
The $X$ propagator is more complicated because of its coupling
to two-meson states. The bare $X$ propagator is constant, 
$iD_{X,0}(p_0,\vec{p})=i/\Delta$, but the full propagator is dressed by 
$D^0$ and $D^{*0}$ meson loops
to all orders (see Fig.~\ref{fig:dress}). 
The bare and full $X$ propagators are indicated by the double dashed
and double lines, respectively. The $D$ mesons are indicated by the 
solid ($D^0$ and $\bar{D}^0$) and dashed  ($\bar{D}^{*0}$ and $D^{*0}$)
lines.
Note that each loop receives contributions of two combinations of $D^{(*)}$
mesons: $D^{0} \bar{D}^{*0}$ and $\bar{D}^0 D^{*0}$ .
Summing the resulting geometric series leads to the full 
$X$ propagator:
\beq
i D_X(p_0,\vec{p}) & = & i D_{X,0}(p_0,\vec{p}) \left[1-D_{X,0}(p_0,\vec{p})
\Sigma(p_0,\vec{p})\right]^{-1}\,, 
\label{dimerprop}
\eeq
where $\Sigma(p)$ is the self energy of the $X$. Using the 
reduced mass of the $D^0$ and $D^{*0}$ mesons
$\mu_X$ and their total mass $M_X=m_{D^0}+m_{D^{*0}}$, 
the self energy can be
written
\beq
\Sigma(p_0,\vec{p}) & = & -2\mu_X g^2 \int {d^3q \over (2\pi)^3} 
\left[{q}^2 -2\mu_X p_0 +\frac{{p}^2}{4}+
\sqrt{1-\frac{4\mu_X}{M_X}}\,\vec{p}\cdot\vec{q} 
- i\epsilon \right]^{-1} \nonumber \\[2pt]
%\nonumber\\
 & = & {2\mu_X g^2 \over 4\pi} \left[ \sqrt{-2\mu_X p_0 
+ {\mu_X \over M_X} {p}^2 - i \epsilon} - {2 \over \pi}\Lambda 
+ {\cal O}(1/\Lambda) \right],
\label{dimerbubble}
\eeq
where the ultraviolet divergence was regulated with a momentum
cutoff $\Lambda$.
Substituting this expression into Eq.~(\ref{dimerprop}) and dropping
terms that vanish as $\Lambda\to \infty$, we obtain the 
full $X$ propagator:
\beq
i D_X(p_0,\vec{p}) & = & {-i4\pi \over 2\mu_Xg^2} 
\left[ -{\gamma} 
+ \sqrt{-2\mu_X p_0 + {\mu_X \over M_X} {p}^2 - i\epsilon} \right]^{-1}\,,
\label{Dprop}
\eeq
where we have matched the bound state pole position
\beq
\gamma \equiv {1\over a}={4\pi\over 2\mu_X} {\Delta \over g^2} 
+ {2 \over \pi}\Lambda\,,
\label{aDD}
\eeq
to the binding momentum of the $X(3872)$: $\gamma=\sqrt{2\mu_X B_X}$
and $a$ is the $D^0 \bar{D}^{*0}$ scattering length. The combination
of bare coupling constants $g^2/\Delta$ must depend
on the cutoff $\Lambda$ as prescribed by Eq.~(\ref{aDD}) since $a$ and
$\gamma$ are physical quantities.

Using the full $X$ propagator we may calculate the scattering
of a $D^0$ or $D^{*0}$ meson (or their antiparticles) off 
the $X(3872)$. 
%%%%%%%%%%%%%%%%%%%%%%%%%%%%%%%%%%%%%%%%%%%%%%%%%%%%%%%%%% 
\begin{figure}[t]
\begin{center}
\includegraphics[width=10cm,angle=0,clip=true]{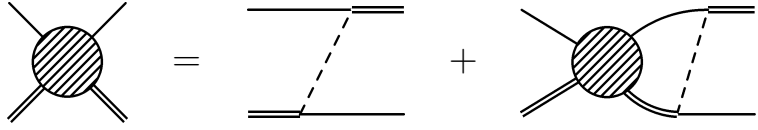}
\end{center}
%\vspace*{-18pt} 
\caption{Integral equation for scattering of a particle $S$ (single
line) off the $X(3872)$ (double line). The dashed line indicates the 
particle $\bar{S}$ complementary to $S$ as explained in the text.}
\label{fig:3bdyeq}
\end{figure}
%%%%%%%%%%%%%%%%%%%%%%%%%%%%%%%%%%%%%%%%%%%%%%%%%%%%%%%%%%%%%%
Because of their different masses, the scattering of a $D^0$ versus
a $D^{*0}$ will lead to a different scattering amplitude
and cross section even though the interaction strength 
$g^2/\Delta$ is the same. 
The scattering amplitude is the solution of the 
integral equation shown in Fig.~\ref{fig:3bdyeq}.
The $X$ and the scattered meson, denoted by $S$,
are represented by a double line and a single line, respectively.
If an $S$ particle is scattered, a second 
complementary particle type,  $\bar{S}$ (represented by a
dashed line in Fig.~\ref{fig:3bdyeq})  appears in the loops.
The masses of the $S$ and
$\bar{S}$ particles are different. For a given scattered particle $S$,
the corresponding particle $\bar{S}$ complementary to $S$
can be read off the flavor wave function
of the $X(3872)$ in Eq.~(\ref{eq:Xflavor}). For example, if $S=D^0$ then
$\bar{S}=\bar{D}^{*0}$.

We now formulate the scattering problem in the center-of-mass frame of the 
$SX$ system. With $k$ the relative momentum of $S$ 
and $X$, the total energy is 
\beq
E = {k^2 \over 2\mu_{SX} } - B_X\,, 
\label{totalEnergy}
\eeq
where $\mu_{SX} = m_S M_X/(M_X+m_S)$ is the reduced mass of the 
$SX$ system and $S=D^0, D^{*0}, \bar{D}^0$, or $\bar{D}^{*0}$. 
The resulting integral equation for the off-shell
$SX$ scattering amplitude is
\beq
T(\vec{k},\vec{p}) & = & {2\pi \gamma \over \mu_X} 
\left(p^2 + k^2 + {2\mu_X \over m_{\bar{S}}} \vec{p}\cdot \vec{k} 
- 2\mu_X E \right)^{-1} \nonumber\\[2pt]
& & + {1 \over (2\pi)^2} \int d\Omega_q \ \int_0^\infty dq\ 
{q^2 \  T(\vec{k},\vec{q}) \over -\gamma + \sqrt{-2\mu_X 
\left(E - q^2/(2\mu_{SX}) \right)- i \epsilon} }  \nonumber\\[2pt]
& & \quad\times \left(p^2 + q^2 + {2\mu_X \over m_{\bar{S}}} 
\vec{p}\cdot \vec{q} - 2\mu_X E \right)^{-1},
\label{TDX}
\eeq
where $\vec k$ and $\vec p$ are the relative momenta in the incoming and outgoing $SX$ system, 
respectively. Performing a partial wave decomposition of  $T(\vec{k},\vec{p})$,
\beq
T(\vec{k},\vec{p}) = \sum_l (2l+1) T_l (k,p) P_l (\cos \theta_{kp})\,,
\label{eq:partialw}
\eeq
where $\theta_{kp}$ is the angle between $\vec{k}$ and $\vec{p}$
and $P_{l}(\cos \theta_{kp})$ is a Legendre polynomial,
and projecting onto the $l$-th partial wave, we obtain
\beq
T_l(k,p) & = & {2\pi \gamma \over\mu_X} {m_{\bar{S}} \over 2\mu_X pk} 
(-1)^l  Q_l\left({m_{\bar{S}} \over 2\mu_X pk} (p^2 + k^2- 2\mu_X E)\right)
\nonumber\\[2pt]
& & + {1 \over \pi} %{1\over 2l+1}
\int_0^\infty dq \ {q^2 \ T_l (k,q) \over -\gamma + 
\sqrt{-2\mu_X \left(E - {q^2/(2\mu_{SX})} \right)- i \epsilon} }  
\nonumber\\[2pt]
& & \quad \times {m_{\bar{S}} \over 2\mu_X pq} 
(-1)^l  Q_l\left({m_{\bar{S}} \over 2\mu_X pq} (p^2 + q^2 - 2\mu_X E )
\right)\,,
\label{TDXl-wave}
\eeq
where
%
%----------------------
\beq
Q_l(z) =\frac{1}{2}\int_{-1}^1 dx \frac{P_l(x)}{z-x} 
\eeq
%----------------------
%
%[hw,l,e]
is a Legendre function of the second kind.
The integral equation for the S-wave amplitude $T_0(k,p)$ reduces to
\beq
T_0(k,p) & = & {2\pi \gamma \over \mu_X} {m_{\bar{S}} \over 4\mu_X pk} 
\ln \left({p^2 + k^2 + {2\mu_X \over m_{\bar{S}}}pk - 2\mu_X E \over 
p^2 + k^2 - {2\mu_X \over m_{\bar{S}}}pk - 2\mu_X E} \right) 
\nonumber\\[2pt]
& & + {1 \over \pi} \int_0^\infty dq \ {q^2 \ T_0(k,q) \over -\gamma 
+ \sqrt{-2\mu_X \left(E - {q^2/(2\mu_{SX})} \right)- i \epsilon} }  
\nonumber\\[2pt]
& & \quad \times {m_{\bar{S}} \over 4\mu_X pq} 
\ln \left({p^2 + q^2 + {2\mu_X \over m_{\bar{S}}}pq - 2\mu_X E \over p^2 + q^2 - {2\mu_X \over m_{\bar{S}}}pq - 2\mu_X E} \right)\,.
\label{TDXS-wave}
\eeq
Solutions of the integral equations (\ref{TDXl-wave}) and (\ref{TDXS-wave})
can be obtained numerically using standard techniques.

The amplitudes $T_l$ are related to the scattering phase shifts through 
the relation:
\beq
T_l(k,k) = {2\pi \over \mu_{SX}}{1 \over k\cot\delta(k)_l - ik}\,.
\label{kcotd}
\eeq
Using the expression for the differential cross section
in terms of the phase shifts:
\beq
{d\sigma\over d\Omega} =
\left| \sum_l {2l+1 \over k\cot\delta_l - ik} P_l(\cos\theta)\right|^2\,,
\label{fL}
\eeq
we obtain  the total cross section for $SX$ scattering:
\beq
\sigma_{XS}(E) = \sum_l {(2l+1)\mu_{SX}^2 \over \pi} 
\left| T_l(k,k) \right|^2\,.
\label{cross-sec}
\eeq

\section{Results and Discussion}
\label{sec:ResDis}

In this section we present our results for the $SX$ scattering amplitude 
and the total cross section at leading order
($S=D^0, D^{*0}, \bar{D}^0$, or $\bar{D}^{*0}$).
With the masses of the $D^0$ and 
$D^{*0}$ mesons fixed, these quantities depend at this order
only on the binding momentum $\gamma$ of the $X(3872)$ (or, equivalently,
the $D^{*0} \bar{D}^0$
scattering length $a=1/\gamma$). Our results are given 
in units of the scattering length and may be scaled
to physical units once $a$ is determined. At present 
the error in the experimental value for $E_X$ in Eq.~(\ref{eq:EX}) implies 
a large error in the scattering length. In particular, 
we obtain the ranges $\gamma=(0 - 36)$ MeV and 
$a=(5.5 - \infty)$ fm with central values  $\gamma=22$ MeV 
and $a=8.8$ fm. 

\begin{figure}[t]
	\centerline{\includegraphics*[width=10cm]{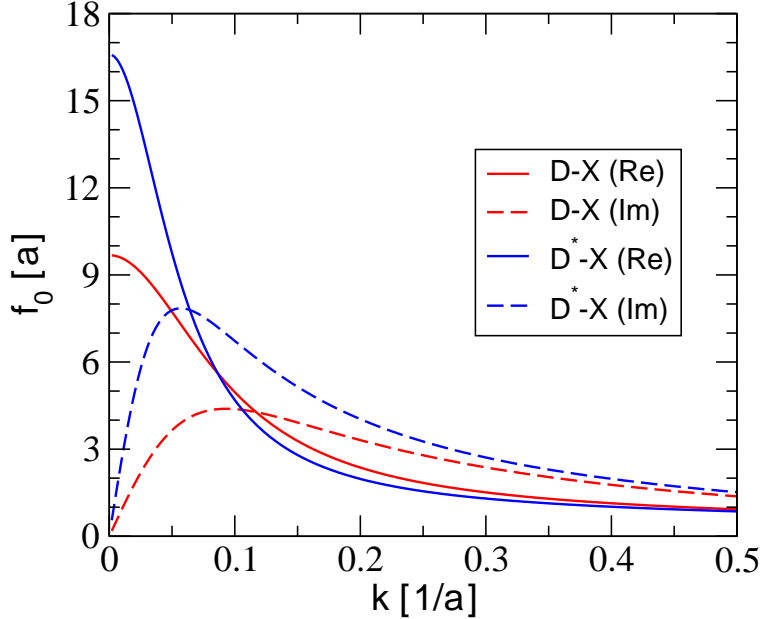}}
	\caption{S-wave scattering amplitude $f_0(k)=1/(k\cot\delta_0(k)-ik)$
                 for scattering of $D^0$ and $D^{*0}$ 
          mesons off the $X(3872)$ in units of the scattering length
          $a$. The scattering amplitude is identical 
          for particles and antiparticles.}
	\label{fig:f0}
\end{figure}

In Fig.~\ref{fig:f0}, we show our results for the S-wave scattering amplitude
$f_0(k)=1/(k\cot\delta_0(k)-ik)$ for the scattering of 
$D^0$ and $D^{*0}$ mesons off the $X(3872)$ for center-of-mass momenta
$k$ from threshold up to $0.5\gamma$. These momenta are still well
below the breakup threshold of $k_B = 1.14 \gamma$ for $D^0 X$ scattering
and $k_B = 1.17 \gamma$ for $D^{0*} X$ scattering.
The scattering amplitude of a particle is the same as that of
its corresponding antiparticle. 
%%%%%%%%%%%%%%%%%%%%%%%%%%%%%%%%%%%%%%%%%%%%%%%%%
%%%%%%%%%%%%%%%%%%%%%%%%%%%%%%%%%%%%%%%%%%%%%%%%%
There is clearly an enhancement of the 
real part of the scattering amplitude at threshold by a factor 
of 10 to 17 depending on whether $D^0 X$ or  $D^{*0}X$ scattering is
considered. This could lead to an enhancement 
of the 
scattering cross section by two orders of magnitude compared to the 
already large cross section for 
$D^0\bar{D}^{*0}$ scattering.
The resulting scattering lengths for $D^0 X$ and  $D^{*0}X$ 
scattering are
\beq
a_{D^0 X}= - 9.7a\,, \quad\mbox{ and }\quad a_{D^{*0} X}= - 16.6\,a\,.
\eeq

\begin{figure}[t]
	\centerline{\includegraphics*[width=10cm]{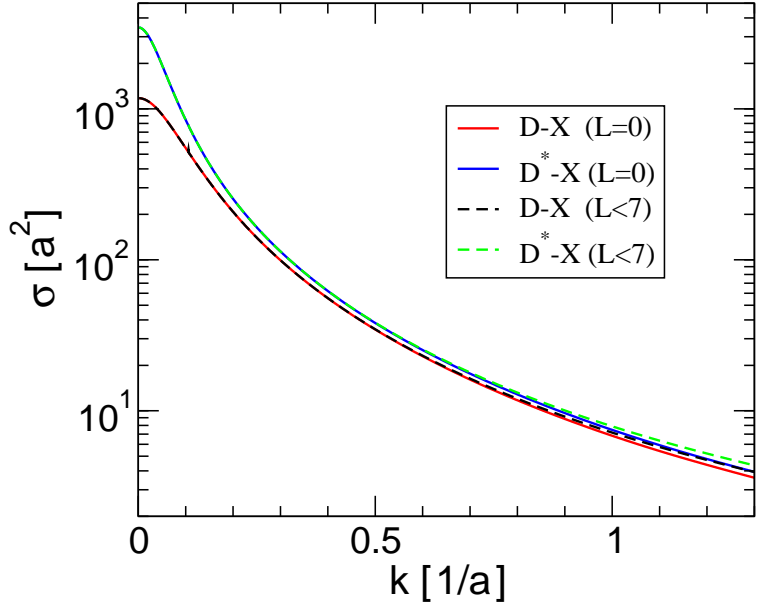}}
	\caption{Total cross section for scattering of $D^0$ and $D^{*0}$ 
          mesons off the $X(3872)$ for S-waves ($l=0$) 
          and including higher partial waves with $l<7$,
          in units of the scattering length $a$.  
          The cross section is the same
          for the scattering of particles as it is for the scattering of 
antiparticles.}
	\label{fig:xsec}
\end{figure}

The corresponding cross sections as a function of the center-of-mass 
momentum $k$ are shown in Fig.~\ref{fig:xsec}. 
The difference between the 
contribution of S-waves ($l=0$) and the full cross section
(including all partial waves up to $l=6$) is negligible 
for momenta below $\gamma$. 
%%%%%%%%%%%%%%%%%%%%%%%%%%%%%%%%%%%%%%%%%%%%%%%%%
%%%%%%%%%%%%%%%%%%%%%%%%%%%%%%%%%%%%%%%%%%%%%%%%%
Using the central value of the scattering length $a$ estimated above, we
obtain for the scale factor $a^2= 0.78$ barn. This factor can become infinite
if the $X(3872)$ is directly at threshold, while the lower bound from the 
error in $E_X$ would give a value of
0.3 barn. Even in this case the total cross section at
threshold will be of the order 300 barns for $D^0 X$ scattering
and 1000 barns for $D^{0*} X$ scattering.

Next we will estimate the leading corrections to our results arising
from the effective range of the $D^0 \bar{D}^{*0}$ system. 
The effective corrections can in principle be calculated up to 
next-to-next-to-leading order in our expansion
in a straightforward way 
\cite{Bedaque:1997qi,Bedaque:1998mb,Griesshammer:2004pe}, but
since the effective range is not known for this system, an estimate
will suffice.

Naively one expects the breakdown scale of the pionless theory
(and the size of the effective range) to be set by
the pion mass just as in nucleon-nucleon scattering
since the longest range interaction not explicitly 
included is the one-pion exchange. 
The situation in the  $D^0 \bar{D}^{*0}$ system is potentially more
interesting
\cite{Fleming:2007rp}
because the mass splitting of the $D^0$ and $D^{*0}$ mesons, 
$\Delta=142$ MeV, is almost
of the same size as the neutral pion mass, $m_\pi=135$ MeV. The range of the 
one-pion exchange interaction is set by the smaller scale
$\mu=\sqrt{\Delta^2-m_\pi^2}\approx 45$ MeV. The mass
difference $\Delta$
appears in the propagator of the exchanged pion because it 
carries energy $q^0=\Delta$, leading to the 
one-pion exchange amplitude for the $D^0 \bar{D}^{*0}$ interaction
\cite{Fleming:2007rp}:
\beq
\frac{g^2}{2f_\pi^2}\frac{\vec{\epsilon}^*\cdot\vec{q}\;  \vec{\epsilon}
\cdot \vec{q}}{\vec{q}^2-\mu^2}\,,
\label{eq:one-pion}
\eeq
where $g$ is the $D$-meson axial transition coupling, $f_\pi$ the pion decay 
constant, $\vec{\epsilon}$ and $\vec{\epsilon}^*$ the polarization
vectors of the incoming and outgoing $D^{0*}$ mesons, and $\vec{q}$ the
three-momentum of the exchanged pion.
However, the work of Ref.~\cite{Fleming:2007rp} shows that,
in part because of the small size of the axial coupling, 
this contribution to NLO effects is very small;
contact corrections will dominate, bringing us back to
an estimate of the effective range of $r_0 \sim 1/m_\pi \sim 1.5$ fm  
($\Lambda_b \sim m_\pi$).
Since the leading corrections to our results are of order $kr_0$ and 
$r_0 \gamma$, we expect the errors to remain less than 20\% even at
momenta close to breakup (using the central value of the binding 
energy of the $X(3872)$
in Eq.~(\ref{eq:EX})). 
For larger momenta, the error is dominated by the 
$kr_0$ correction and will increase to 35\% at momenta of order
$45$ MeV. 
Compared to the errors from effective range corrections
the effects from the charged $D$ meson channel can safely be neglected;
they only enter at  much higher momenta since
the energy difference between the neutral and charged thresholds of about 
8 MeV corresponds to typical momenta of order 130 MeV or $6\gamma$.

To observe the three body interactions described here requires identifying an
experimental process where, for example, two $D^0$ mesons
and one $\bar{D}^{*0}$ are 
produced very near each other in space and time.  One possibility is
provided by 
the decay of the $B_c$ particle.  The $B_c$ was discovered through its
decays into $J/\psi$ in Run I at CDF \cite{Abe:1998wi,Abe:1998fb}.
Particle Data Book (2007) averages are: $m_{B_c} = (6.286 \pm 0.005)$ GeV and
$\tau_{B_c} = (0.46 \pm 0.07) \times 10^{-12}$ s. Several analyses have
been undertaken (see references in \cite{Brambilla:2004wf}, Chapter 4)
to determine the most likely mode by
which the $B_c$ would decay; the $b$ quark decaying first, the $c$
quark decaying first, the two valence quarks annihilating, etc.
%%%%%%%%%%%%%%%%%%%%%%%%%%%%%%%%%%%%%%%%%%%%%%%%%
\begin{figure}[t]
\centerline{\includegraphics*[width=6cm]{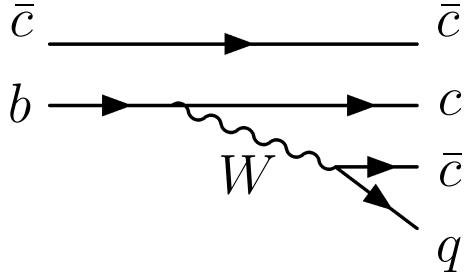}}
\caption{\label{fig:quarkdecay}An example of a quark-level $B_c$ decay
yielding three charmed/anticharmed quarks in the final state}
\end{figure}
%%%%%%%%%%%%%%%%%%%%%%%%%%%%%%%%%%%%%%%%%%%%%%%%%
For access to the three body neutral $D^{(*)}$ meson interactions, we
require that the $B_c$ decay in a mode such as that in
Fig.~\ref{fig:quarkdecay}, yielding 
three $c (\overline c)$ quarks in the final state.  The mass total for
the three body $D$ meson system will be 5.75 to 5.88 GeV (depending
upon whether the third $D$ is a $D^0$ or a $D^{*0}$).  Along with
the additional meson in a P-wave required to balance the $B_c$ charge
and spin, there is
not much phase space available.  The $q$ in the diagram could be the
Cabbibo-favored strange quark or the Cabbibo-suppressed ($V_{cd}/V_{cs}
\sim$ 1/4) down quark.  Relative suppression of both decay modes is
caused by Pauli interference between the spectator $\overline c$ and
the $\overline c$ from the (second vertex of the) weak decay of the
$b$ quark.  From Ref.~\cite{Brambilla:2004wf}: the estimate for the
quark level $B_c \rightarrow \overline c c \overline c s$ decay is
about 1.4\% \cite{Kiselev:2002vz}; the detection efficiency for a
single $D^0$ is expected to be 11 to 31\%; the production cross
section at the LHC (not including feeddowns, which may provide an
increase of more than a factor of five, but might also be harder to
identify) of the $B_c$ is expected at the 30-60 nb level.  At LHCb,
the yield will be perhaps 10$^7$ $B_c$ events per week of running.  So
the prospect of seeing the three body neutral $D^{(*)}$ meson interactions
through $B_c$ decay may well be difficult, but is worth investigating.

Another possibility for observing $XD^{(*)}$ scattering would be in a 
$B \bar{B}$ production event where one $B$ decays to an $X(3872)$ and
its partner $B$ decays to a $D^{(*)}$ \cite{Ericsuggest}.
Conditions at asymmetric $B$-factories do
not favor the interaction of the decay products \cite{Olsen}, but
conditions at the LHC may.  
Heavy flavor production at the LHC has received extensive attention because
of the need to correct for Standard Model background processes in the
search for the Higgs or new physics.  The theoretical and experimental
prospects are reviewed 
in Ref.~\cite{Alekhin:2005dy}.  Production of
the $X(3872)$ under CDF conditions is already 
dominated by prompt events \cite{lifetime}, and heavy flavor 
production at the
LHC is expected to be dominated by $gg$ fusion.  
The cross section for $b \bar b$ at the LHC is about 0.5 mb.
Predictions of $b \bar b$ correlations indicate
that there are events where the opening angle between 
them may be small (see Fig. 5 on pg. 266 in 
Ref.~\cite{Alekhin:2005dy});
the $gg \rightarrow Q \bar Q$ cross section is dominated by rapidity 
differences less than one \cite{Ellis:1991qj}.  
This is important if we expect their
decay products to interact. 
The cross section for $c\bar c$ production at the LHC is larger 
(10 mb), but a minimum of $c \bar c c \bar c$ would be required
to produce an $XD^{(*)}$ scattering event.
The ALICE detector will be sensitive to quarkonia-type
particles such
as the $X(3872)$, as well as $D$ particles, which might
be produced promptly, 
while the LHCb detector is optimized to look at $B$ decay
products at larger rapidity, where an $XD^{(*)}$ interaction might
be more likely.    Another process that might yield
smaller opening angles and an enhanced opportunity
for final state interactions would be the production of $X$ and $D^{(*)}$
from $b \bar b b \bar b$, which is expected at the LHC at a 
cross section of  $\sim$ 500 fb \cite{Reiter:2009dk}.  
Nucleus-nucleus collisions in the LHC
will also produce $X(3872)$'s along with associated $D^{(*)}$ mesons.

The effect of final state interactions involving the $X$ and $D^{(*)}$, 
along with the characteristics
of the two-body resonance, may be reflected in the distribution in
space and energy of the detected particles.  Typically, (e.g., 
Ref.~\cite{Amado:1993xz}) there will be an enhancement or de-enhancement
of the total cross section relative to the situation where three
body scattering does not occur in the final state.  In particular,
the behavior of the $X(3872)$ produced in isolation should be
distinguishable from its behavior when in the presence of 
$S=D^0, D^{*0}, \bar{D}^0$, or $\bar{D}^{*0}$, a situation that may
be accessible in the rich environment of the LHC.  
For example, if to leading order we can assume that the $X$ and
$D^{(*)}$ are produced isotropically in their opening angle, we might
attribute violations of isotropy at small opening angles to the effect of
the large $XD^{(*)}$ scattering in the final state.

\begin{acknowledgments}

We thank Eric Braaten, Bira van Kolck, and Thomas Mehen for useful discussions.
This research was supported in part by the DFG through
SFB/TR 16 \lq\lq Subnuclear structure of matter,'' the BMBF
under contract No. 06BN411, and the US Department of Energy under
DE-FG02-05ER41368.

\end{acknowledgments}

%%%%%%%%%%%%%%%%%%%%%%%%%%%%%%%%%%%%%%%%%%%%%%%%
%Bibliography
%%%%%%%%%%%%%%%%%%%%%%%%%%%%%%%%%%%%%%%%%%%%%%%%


\begin{thebibliography}{99}
%\cite{Palano:2009zz}
\bibitem{Palano:2009zz}
  A.~Palano  [BaBar Collaboration],
  %``Charm And Charmonium Spectroscopy At Babar,''
  Int.\ J.\ Mod.\ Phys.\  A {\bf 24}, 343 (2009).
  %%CITATION = IMPAE,A24,343;%%

%\cite{Bracko:2009zz}
\bibitem{Bracko:2009zz}
  M.~Bracko  [BELLE Collaboration],
  %``Charm And Charmonium Spectroscopy At Belle,''
  Int.\ J.\ Mod.\ Phys.\  A {\bf 24}, 334 (2009).
  %%CITATION = IMPAE,A24,334;%%

%\cite{Mendez:2007ey}
\bibitem{Mendez:2007ey}
  H.~Mendez  [CLEO Collaboration],
  %``Recent CLEO-c results,''
  AIP Conf.\ Proc.\  {\bf 917}, 338 (2007)
  [arXiv:hep-ex/0702008].
  %%CITATION = APCPC,917,338;%%

%\cite{Nielsen:2008tt}
\bibitem{Nielsen:2008tt}
  M.~Nielsen,
  %``Review and interpretation of the new heavy states discovered at the $B^-$
  %factories,''
  arXiv:0810.5526 [hep-ph].
  %%CITATION = ARXIV:0810.5526;%%

%\cite{Robutti:2009af}
\bibitem{Robutti:2009af}
  E.~Robutti  [BaBar Collaboration],
  %``Quarkonium States at $B^-$ Factories,''
  arXiv:0903.0450 [hep-ex].
  %%CITATION = ARXIV:0903.0450;%%


%\cite{Swanson:2006st}
\bibitem{Swanson:2006st}
  E.~S.~Swanson,
  %``The new heavy mesons: A status report,''
  Phys.\ Rept.\  {\bf 429}, 243 (2006)
  [arXiv:hep-ph/0601110].
  %%CITATION = PRPLC,429,243;%%

%\cite{Eichten:2007qx}
\bibitem{Eichten:2007qx}
  E.~Eichten, S.~Godfrey, H.~Mahlke and J.~L.~Rosner,
  %``Quarkonia and their transitions,''
  Rev.\ Mod.\ Phys.\  {\bf 80}, 1161 (2008)
  [arXiv:hep-ph/0701208].
  %%CITATION = RMPHA,80,1161;%%

%\cite{Voloshin:2007dx}
\bibitem{Voloshin:2007dx}
  M.~B.~Voloshin,
  %``Charmonium,''
  Prog.\ Part.\ Nucl.\ Phys.\  {\bf 61}, 455 (2008)
  [arXiv:0711.4556 [hep-ph]].
  %%CITATION = PPNPD,61,455;%%

%\cite{Godfrey:2008nc}
\bibitem{Godfrey:2008nc}
  S.~Godfrey and S.~L.~Olsen,
  %``The Exotic XYZ Charmonium-like Mesons,''
  arXiv:0801.3867 [hep-ph].
  %%CITATION = ARXIV:0801.3867;%%

\bibitem{Lange}
J.S.~Lange 
% ``PANDA-Experiment with Antiprotons at FAIR,''
% http://www-conf.kek.jp/qwg08/
{\it International Workshop on Heavy Quarkonia}, 
December 2-5, 2008, Nara, Japan.

%\cite{Voloshin:1976ap}
\bibitem{Voloshin:1976ap}
  M.~B.~Voloshin and L.~B.~Okun,
  %``Hadron Molecules And Charmonium Atom,''
  JETP Lett.\  {\bf 23}, 333 (1976)
  [Pisma Zh.\ Eksp.\ Teor.\ Fiz.\  {\bf 23}, 369 (1976)].
  %%CITATION = ZFPRA,23,369;%%

%\cite{De Rujula:1976qd}
\bibitem{De Rujula:1976qd}
  A.~De Rujula, H.~Georgi and S.~L.~Glashow,
  %``Molecular Charmonium: A New Spectroscopy?,''
  Phys.\ Rev.\ Lett.\  {\bf 38}, 317 (1977).
  %%CITATION = PRLTA,38,317;%%

%\cite{Tornqvist:1991ks}
\bibitem{Tornqvist:1991ks}
  N.~A.~Tornqvist,
  %``Possible large deuteron - like meson meson states bound by pions,''
  Phys.\ Rev.\ Lett.\  {\bf 67}, 556 (1991).
  %%CITATION = PRLTA,67,556;%%


%\cite{Choi:2003ue}
\bibitem{Choi:2003ue}
  S.~K.~Choi {\it et al.}  [Belle Collaboration],
  %``Observation of a new narrow charmonium state in exclusive B+- --> K+-  pi+
  %pi- J/psi decays,''
  Phys.\ Rev.\ Lett.\  {\bf 91}, 262001 (2003)
  [arXiv:hep-ex/0309032].
  %%CITATION = PRLTA,91,262001;%%


%\cite{Acosta:2003zx}
\bibitem{Acosta:2003zx}
  D.~E.~Acosta {\it et al.}  [CDF II Collaboration],
  %``Observation of the narrow state $X(3872) \to J/\psi \pi^+ \pi^-$ in
  %$\bar{p}p$  collisions at $\sqrt{s} = 1.96$ TeV,''
  Phys.\ Rev.\ Lett.\  {\bf 93}, 072001 (2004)
  [arXiv:hep-ex/0312021].
  %%CITATION = PRLTA,93,072001;%%

%\cite{Abazov:2004kp}
\bibitem{Abazov:2004kp}
  V.~M.~Abazov {\it et al.}  [D0 Collaboration],
  %``Observation and properties of the $X(3872)$ decaying to $J/\psi \pi^+
  %\pi^-$ in $p\bar{p}$ collisions at $\sqrt{s} = 1.96$ TeV,''
  Phys.\ Rev.\ Lett.\  {\bf 93}, 162002 (2004)
  [arXiv:hep-ex/0405004].
  %%CITATION = PRLTA,93,162002;%%

%\cite{Aubert:2004ns}
\bibitem{Aubert:2004ns}
  B.~Aubert {\it et al.}  [BABAR Collaboration],
  %``Study of the $B \to J/\psi K^- \pi^+ \pi^-$ decay and measurement of the $B
  %\to X(3872) K^-$ branching fraction,''
  Phys.\ Rev.\  D {\bf 71}, 071103 (2005)
  [arXiv:hep-ex/0406022].
  %%CITATION = PHRVA,D71,071103;%%

%\cite{Close:2003sg}
\bibitem{Close:2003sg}
  F.~E.~Close and P.~R.~Page,
  %``The D*0 D0bar threshold resonance,''  [tests]
  Phys.\ Lett.\  B {\bf 578}, 119 (2004)
  [arXiv:hep-ph/0309253].
  %%CITATION = PHLTA,B578,119;%%

%\cite{Pakvasa:2003ea}
\bibitem{Pakvasa:2003ea}
  S.~Pakvasa and M.~Suzuki,
  %``On the hidden charm state at 3872 MeV,'' [qm tests]
  Phys.\ Lett.\  B {\bf 579}, 67 (2004)
  [arXiv:hep-ph/0309294].
  %%CITATION = PHLTA,B579,67;%%


%\cite{Wong:2003xk}
\bibitem{Wong:2003xk}
  C.~Y.~Wong,
  %``Molecular States of Heavy Quark Mesons,''
  Phys.\ Rev.\  C {\bf 69}, 055202 (2004)
  [arXiv:hep-ph/0311088].
  %%CITATION = PHRVA,C69,055202;%%


%\cite{Braaten:2008nv}
\bibitem{Braaten:2008nv}
  E.~Braaten,
  %``Exotic c c-bar Mesons,''
  arXiv:0808.2948 [hep-ph].
  %%CITATION = ARXIV:0808.2948;%%


%\cite{Braaten:2007dw}
\bibitem{Braaten:2007dw}
  E.~Braaten and M.~Lu,
  %``Line Shapes of the X(3872),''
  Phys.\ Rev.\  D {\bf 76}, 094028 (2007)
  [arXiv:0709.2697 [hep-ph]].
  %%CITATION = PHRVA,D76,094028;%%

%\cite{Braaten:2007ft}
\bibitem{Braaten:2007ft}
  E.~Braaten and M.~Lu,
  %``The Effects of Charged Charm Mesons on the Line Shapes of the X(3872),''
  Phys.\ Rev.\  D {\bf 77}, 014029 (2008)
  [arXiv:0710.5482 [hep-ph]].
  %%CITATION = PHRVA,D77,014029;%%

%\cite{:2008su}
\bibitem{:2008su}
  I.~Adachi {\it et al.}  [Belle Collaboration],
  %``Study of the $B \to X(3872)(D*^0 \bar{D}^0) K$ decay,''
  arXiv:0810.0358 [hep-ex].
  %%CITATION = ARXIV:0810.0358;%%

%\cite{Hanhart:2007yq}
\bibitem{Hanhart:2007yq}
  C.~Hanhart, Yu.~S.~Kalashnikova, A.~E.~Kudryavtsev and A.~V.~Nefediev,
  %``Reconciling the X(3872) with the near-threshold enhancement in the
  %D^0\bar{D}^{*0} final state,''
  Phys.\ Rev.\  D {\bf 76}, 034007 (2007)
  [arXiv:0704.0605 [hep-ph]].
  %%CITATION = PHRVA,D76,034007;%%

%\cite{Belle:2008te}
\bibitem{Belle:2008te}
 I.~Adachi {\it et al.}  [Belle Collaboration],
%  ``Study of $X(3872)$ in $B$ meson decays,''
  arXiv:0809.1224 [hep-ex].

%\cite{Aubert:2008gu}
\bibitem{Aubert:2008gu}
  B.~Aubert {\it et al.}  [BABAR Collaboration],
%  ``A study of $B \to X(3872) K$, with $X(3872) \to J/\psi \pi^+ \pi^-$,''
  Phys.\ Rev.\  D {\bf 77}, 111101 (2008)
  [arXiv:0803.2838 [hep-ex]].

\bibitem{CDF-QWG08}
CDF collaboration, talk presented by Thomas~Kuhr
at the QWG08 meeting, December 2-5 2008, Nara, Japan 
(see http://www-conf.kek.jp/qwg08/ ).

\bibitem{Braaten:eft09}
Eric Braaten,
talk presented at the EFT09 meeting, February 1-6, 2009, Valencia, Spain
(see http://ific.uv.es/eft09/ ).

%\cite{Cawlfield:2007dw}
\bibitem{Cawlfield:2007dw}
  C.~Cawlfield {\it et al.}  [CLEO Collaboration],
  %``A precision determination of the D0 mass,''
  Phys.\ Rev.\ Lett.\  {\bf 98}, 092002 (2007)
  [arXiv:hep-ex/0701016].
  %%CITATION = PRLTA,98,092002;%%


%\cite{Bugg:2004rk}
\bibitem{Bugg:2004rk}
  D.~V.~Bugg,
  %``Reinterpreting several narrow `resonances' as threshold cusps,''
  Phys.\ Lett.\  B {\bf 598}, 8 (2004)
  [arXiv:hep-ph/0406293].
  %%CITATION = PHLTA,B598,8;%%



%\cite{Voloshin:2007hh}
\bibitem{Voloshin:2007hh}
  M.~B.~Voloshin,
  %``Isospin properties of the X state near the D {\bar D}^{*} threshold,''
  Phys.\ Rev.\  D {\bf 76}, 014007 (2007)
  [arXiv:0704.3029 [hep-ph]].
  %%CITATION = PHRVA,D76,014007;%%


\bibitem{Braaten:2004rn}
  E.~Braaten and H.-W.~Hammer,
  %``Universality in Few-body Systems with Large Scattering Length,''
  Phys.\ Rept.\  {\bf 428}, 259 (2006)
  [arXiv:cond-mat/0410417].
  %%CITATION = PRPLC,428,259;%%

%\cite{vanKolck:1999mw}
\bibitem{vanKolck:1999mw}
  U.~van Kolck,
  %``Effective field theory of nuclear forces,''
  Prog.\ Part.\ Nucl.\ Phys.\  {\bf 43}, 337 (1999)
  [arXiv:nucl-th/9902015].
  %%CITATION = PPNPD,43,337;%%

%\cite{Beane:2000fx}
\bibitem{Beane:2000fx}
  S.~R.~Beane, P.~F.~Bedaque, W.~C.~Haxton, D.~R.~Phillips and M.~J.~Savage,
  %``From hadrons to nuclei: Crossing the border,''
  arXiv:nucl-th/0008064.
  %%CITATION = NUCL-TH/0008064;%%


%\cite{Bedaque:2002mn}
\bibitem{Bedaque:2002mn}
  P.~F.~Bedaque and U.~van Kolck,
  %``Effective field theory for few-nucleon systems,''
  Ann.\ Rev.\ Nucl.\ Part.\ Sci.\  {\bf 52}, 339 (2002)
  [arXiv:nucl-th/0203055].
  %%CITATION = ARNUA,52,339;%%

%\cite{Epelbaum:2008ga}
\bibitem{Epelbaum:2008ga}
  E.~Epelbaum, H.-W.~Hammer and U.-G.~Mei\ss ner,
  %``Modern Theory of Nuclear Forces,''
  arXiv:0811.1338 [nucl-th], to appear in Rev.\ Mod.\ Phys.\ (2009).
  %%CITATION = ARXIV:0811.1338;%%


%\cite{Fleming:2007rp}
\bibitem{Fleming:2007rp}
  S.~Fleming, M.~Kusunoki, T.~Mehen and U.~van Kolck,
  %``Pion interactions in the $X(3872)$,''
  Phys.\ Rev.\  D {\bf 76}, 034006 (2007)
  [arXiv:hep-ph/0703168].
  %%CITATION = PHRVA,D76,034006;%%

%\cite{Fleming:2008yn}
\bibitem{Fleming:2008yn}
  S.~Fleming and T.~Mehen,
  %``Hadronic Decays of the X(3872) to chi_{cJ} in Effective Field Theory,''
  Phys.\ Rev.\  D {\bf 78}, 094019 (2008)
  [arXiv:0807.2674 [hep-ph]].
  %%CITATION = PHRVA,D78,094019;%%

%\cite{Braaten:2003he}
\bibitem{Braaten:2003he}
  E.~Braaten and M.~Kusunoki,
  %``Low-energy universality and the new charmonium resonance at 3870-MeV,''
  Phys.\ Rev.\  D {\bf 69}, 074005 (2004)
  [arXiv:hep-ph/0311147].
  %%CITATION = PHRVA,D69,074005;%%

%\cite{AlFiky:2005jd}
\bibitem{AlFiky:2005jd}
  M.~T.~AlFiky, F.~Gabbiani and A.~A.~Petrov,
  %``X(3872): Hadronic Molecules in Effective Field Theory,'' [DD bound state
% EFT integ out pi]
  Phys.\ Lett.\  B {\bf 640}, 238 (2006)
  [arXiv:hep-ph/0506141].
  %%CITATION = PHLTA,B640,238;%%

\bibitem{Xproduction}
%\cite{Braaten:2004rw}
%\bibitem{Braaten:2004rw}
  E.~Braaten and M.~Kusunoki,
  %``Production of the X(3870) at the Upsilon(4S) by the coalescence of  charm
  %mesons from B decays,''
  Phys.\ Rev.\  D {\bf 69}, 114012 (2004)
  [arXiv:hep-ph/0402177];
  %%CITATION = PHRVA,D69,114012;%%
%\cite{Braaten:2004fk}
%\bibitem{Braaten:2004fk}
  E.~Braaten, M.~Kusunoki and S.~Nussinov,
  %``Production of the X(3870) in B meson decay by the coalescence of charm
  %mesons,''
  Phys.\ Rev.\ Lett.\  {\bf 93}, 162001 (2004)
  [arXiv:hep-ph/0404161];
  %%CITATION = PRLTA,93,162001;%%
%\cite{Braaten:2004jg}
%\bibitem{Braaten:2004jg}
  E.~Braaten,
  %``Inclusive production of the X(3872),''
  Phys.\ Rev.\  D {\bf 73}, 011501 (2006)
  [arXiv:hep-ph/0408230];
  %%CITATION = PHRVA,D73,011501;%%
%\cite{Braaten:2004ai}
%\bibitem{Braaten:2004ai}
  E.~Braaten and M.~Kusunoki,
  %``Exclusive production of the X(3872) in B meson decay,''
  Phys.\ Rev.\  D {\bf 71}, 074005 (2005)
  [arXiv:hep-ph/0412268].
  %%CITATION = PHRVA,D71,074005;%%

\bibitem{Xdecay}
%\cite{Braaten:2005jj}
%\bibitem{Braaten:2005jj}
  E.~Braaten and M.~Kusunoki,
  %``Factorization in the production and decay of the X(3872),''
  Phys.\ Rev.\  D {\bf 72}, 014012 (2005)
  [arXiv:hep-ph/0506087];
  %%CITATION = PHRVA,D72,014012;%%
%\cite{Braaten:2005ai}
%\bibitem{Braaten:2005ai}
  E.~Braaten and M.~Kusunoki,
  %``Decays of the X(3872) into J/psi and light hadrons,''
  Phys.\ Rev.\  D {\bf 72}, 054022 (2005)
  [arXiv:hep-ph/0507163];
  %%CITATION = PHRVA,D72,054022;%%
%\cite{Braaten:2006sy}
%\bibitem{Braaten:2006sy}
  E.~Braaten and M.~Lu,
  %``Operator product expansion in the production and decay of the X(3872),''
  Phys.\ Rev.\  D {\bf 74}, 054020 (2006)
  [arXiv:hep-ph/0606115];
  %%CITATION = PHRVA,D74,054020;%%
%\cite{Braaten:2007sh}
%\bibitem{Braaten:2007sh}
  E.~Braaten,
  %``An Estimate of the Partial Width for X(3872) into p p-bar,''
  Phys.\ Rev.\  D {\bf 77}, 034019 (2008)
  [arXiv:0711.1854 [hep-ph]].
  %%CITATION = PHRVA,D77,034019;%%

\bibitem{Efimov-70}
V.~Efimov, 
%"Energy Levels Arising from Resonant Two-body Forces in a Three-body System," 
Phy. Lett. {\bf 33B}, 563 (1970).

%\cite{Bedaque:1997qi}
\bibitem{Bedaque:1997qi}
  P.~F.~Bedaque and U.~van Kolck,
  %``Nucleon deuteron scattering from an effective field theory,''
  Phys.\ Lett.\  B {\bf 428}, 221 (1998)
  [arXiv:nucl-th/9710073].
  %%CITATION = PHLTA,B428,221;%%


%\cite{Bedaque:1998mb}
\bibitem{Bedaque:1998mb}
  P.~F.~Bedaque, H.-W.~Hammer and U.~van Kolck,
  %``Effective Theory for Neutron-Deuteron Scattering: Energy Dependence,''
  Phys.\ Rev.\  C {\bf 58}, 641 (1998)
  [arXiv:nucl-th/9802057].
  %%CITATION = PHRVA,C58,R641;%%

%\cite{Griesshammer:2004pe}
\bibitem{Griesshammer:2004pe}
  H.~W.~Griesshammer,
  %``Improved Convergence in the Three-Nucleon System at Very Low Energies,''
  Nucl.\ Phys.\  A {\bf 744}, 192 (2004)
  [arXiv:nucl-th/0404073].
  %%CITATION = NUPHA,A744,192;%%

%\cite{Abe:1998wi}
\bibitem{Abe:1998wi}
  F.~Abe {\it et al.}  [CDF Collaboration],
  %``Observation of the $B_c$ meson in $p\bar{p}$ collisions at $\sqrt{s} =
  %1.8$ TeV,''
  Phys.\ Rev.\ Lett.\  {\bf 81}, 2432 (1998)
  [arXiv:hep-ex/9805034].
  %%CITATION = PRLTA,81,2432;%%

%\cite{Abe:1998fb}
\bibitem{Abe:1998fb}
  F.~Abe {\it et al.}  [CDF Collaboration],
  %``Observation of $B_c$ mesons in $p\bar{p}$ collisions at $\sqrt{s} = 1.8$
  %TeV,''
  Phys.\ Rev.\  D {\bf 58}, 112004 (1998)
  [arXiv:hep-ex/9804014].
  %%CITATION = PHRVA,D58,112004;%%

%\cite{Brambilla:2004wf}
\bibitem{Brambilla:2004wf}
  N.~Brambilla {\it et al.}  [Quarkonium Working Group],
  %``Heavy quarkonium physics,''
  arXiv:hep-ph/0412158.
  %%CITATION = HEP-PH/0412158;%%

%\cite{Kiselev:2002vz}
\bibitem{Kiselev:2002vz}
  V.~V.~Kiselev,
  %``Exclusive decays and lifetime of B/c meson in QCD sum rules. ((U)),''
  arXiv:hep-ph/0211021.
  %%CITATION = HEP-PH/0211021;%%

\bibitem{Ericsuggest}We thank E. Braaten for 
suggesting this possibility.

\bibitem{Olsen}
S.~L.~Olsen, private communication.

%\cite{Alekhin:2005dy}
\bibitem{Alekhin:2005dy}
  S.~Alekhin {\it et al.},
  %``HERA and the LHC - A workshop on the implications of HERA for LHC  physics:
  %Proceedings Part B,''
  arXiv:hep-ph/0601013.
  %%CITATION = HEP-PH/0601013;%%

\bibitem{lifetime}
CDF II Collaboration, {\sl The ``Lifetime'' Distribution of $X(3872)$ Mesons
Produced in $p \bar p$ Collisions at CDF}, CDF Note 7159, August 3, 2004.


%\cite{Ellis:1991qj}
\bibitem{Ellis:1991qj}
  R.~K.~Ellis, W.~J.~Stirling and B.~R.~Webber,
  %``QCD and collider physics,''
  Camb.\ Monogr.\ Part.\ Phys.\ Nucl.\ Phys.\ Cosmol.\  {\bf 8}, 1 (1996).
  %%CITATION = CMPCE,8,1;%%

%\cite{Reiter:2009dk}
\bibitem{Reiter:2009dk}
  T.~Reiter,
  %``An Automated Approach for $q\bar{q}\to b\bar{b}b\bar{b}$ at Next-to-Leading
  %Order QCD,''
  arXiv:0903.4648 [hep-ph].
  %%CITATION = ARXIV:0903.4648;%%


%\cite{Amado:1993xz}
\bibitem{Amado:1993xz}
  R.~D.~Amado and J.~V.~Noble,
  %``Final-state interaction effects in weak three-body decays,''
  Phys.\ Rev.\  {\bf 185}, 1993 (1969).
  %%CITATION = PHRVA,185,1993;%%


\end{thebibliography}
\end{document}